\documentclass[twocolumn]{article}
\usepackage[margin = 0.75in]{geometry}

\usepackage{graphicx}
\usepackage{amssymb}
\usepackage{amsmath}
\usepackage{amsthm}
\usepackage{epstopdf}
\usepackage{enumerate}

\usepackage[font={small}]{caption}

\usepackage{algorithm}
\usepackage[noend]{algpseudocode}

\DeclareMathOperator{\thh}{^\text{th}}

\DeclareMathOperator{\var}{var}

\newcommand{\E}{\ensuremath{\mathbb{E}}}

\theoremstyle{definition}

\newtheorem{remark}{Remark}

\newtheorem{prop}{Proposition}
\newtheorem*{example}{Example}

\title{Time-Series Estimation from Randomly Time-Warped Observations}
 \author{\.{I}lker Bayram\footnote{The author was with Analog Devices Inc, Analog Garage, Boston, MA, at the time of submission.}\\
ibayram@ieee.org}
\date{}
\begin{document}
\maketitle

\begin{abstract}
We consider the problem of estimating a signal from its warped observations. Such estimation is commonly performed by altering the observations through some inverse-warping, or solving a computationally demanding optimization formulation. While these may be unavoidable if observations are few, when large amounts of warped observations are available, the cost of running such algorithms can be prohibitive. We consider the scenario where we have many observations, and propose a computationally simple algorithm for estimating the function of interest. We demonstrate the utility of the algorithm on streaming biomedical signals.
\end{abstract}

\section{Introduction}

Suppose we are interested in a function $f(t)$, but we make time-warped observations of the form 
\begin{equation}\label{eqn:f}
f_k(t) = f\bigl( g_k(t) \bigr) + n_k(t),
\end{equation}
where $g_k(\cdot)$ is an unknown monotonic function, warping the time variable, and $n_k(\cdot)$ denotes a noise process. An instance of $f$ and $f_k$'s are shown in Fig.~\ref{fig:observations}.  Also, Fig.~\ref{fig:warp} shows an instance of $g_k$, compared to the identity mapping. This letter proposes a simple method for recovering $f$ given only $f_k$'s. Our main running assumption is that $g_k$ deviates around the identity mapping.

This problem is of interest in applications that involve pseudo-periodic stochastic signals. For instance, in an application like radar based vital signs monitoring \cite{chen}, the signal obtained via radar contains information about the heart activity, but the shape of this signal is unknown, because it ultimately depends on the radar monitoring conditions. The knowledge of the average waveform would be valuable in such an estimation scenario. Another application involves PPG signals. PPG waveform carries information about the condition of the subject \cite{elg12p14, all07}. Therefore, it is of interest to estimate a `typical' waveform associated with an individual over a course of time. But this is hindered by the time-variation of the waveforms. The current letter addresses this scenario (see, specifically Section~\ref{sec:PPG}).

A common approach to recovering $f$ involves inverse warping $f_k$'s \cite{FDA, kur11nips, Tucker2014, jam07p480, ram98p351}. However, a direct implementation of this idea quickly runs into a first obstacle~: when warping $f_k$ to closely match $f$, we need to know $f$. To avoid this problem, the works cited above resort to workarounds, which incur a computational cost. This cost may be acceptable when only a few $f_k$'s are available. However, when many $f_k$'s are at hand, the computational cost becomes prohibitive. 

The method proposed in this letter consciously avoids time-warping. Our aim is to introduce an algorithm that scales well with the number of warped observations. The method interprets each $f_k$ as a point in a bounded manifold, and seeks the `center' of the manifold by resorting to the notion of graph centrality \cite{Newman}. We take this center to be an approximation of the underlying $f(\cdot)$. The proposed algorithm consists of a few interpretable steps, in contrast to a black box approach like a deep neural network.

Our motivation to avoid estimating the warping $g_k$'s stems from the complexity and unreliability of recovering time-warps. In our experience, time-(de)warping occasionally requires fine tuning, and this can be an obstacle to automation. Therefore, even though estimating $g_k$'s might be  necessary in many situations, it is desirable to understand when it can be avoided. 
We demonstrate in this letter that estimating $\gamma_k$'s can be avoided when
\begin{itemize}
\item The number of observations $K$ is large,
\item $g_k(t) \approx t$, where the relation holds \emph{sufficiently} well. 
\end{itemize}

\subsection*{Related Work}
The literature on time-warping, or curve alignment is rich. We review only some of the prominent approaches that are relevant for the problem of interest for this letter.

Kneip and Gasser \cite{knei92p266} use structural functionals that extract landmarks identified on the signals. They then recover the warping (or local shift functions) that align the landmarks. In recovering the local shift functions, they rely on the an average location for each landmark point.

In \cite{ram98p351}, Ramsay and Li propose an iterative algorithm to align $f_k$'s. Using the `cross-sectional average' $f^0(\cdot) = \frac{1}{N}\sum_{k=1}^N\, f_k(\cdot)$, they solve a regularized minimization problem involving (inverse) warp functions $w_k$ so that $f_k(w_k(\cdot))$ is closer to $f^0(\cdot)$. Then they update the cross sectional average using $f_k(w_k(\cdot))$'s and iterate this procedure. 

James \cite{jam07p480} proposes a minimization formulation based on `moments'. In this context, moments are defined to extract the properties of a curve that vary with domain variations due to warping. The minimization formulation seeks to find (inverse) warpings $w_k$ such that the moments of $f_k(w_k(\cdot))$ are all similar, and $f_k(w_k(\cdot))$ is not too far from $f_k(\cdot)$.

In \cite{kur11nips}, Kurtek et al. consider the equivalence classes obtained by warping an associated square root velocity function of each $f_k$. They compute a `mean' of the observed equivalence classes, find the center of the mean class, and then warp each $f_k$ so as to approximate this center as well as possible to align $f_k$'s. Then, $f$ is estimated by averaging the inverse warped functions. We also refer to \cite{tuc13p50, Tucker2014} in this context.

Zhou and De la Torrre  \cite{zho16p279} combine canonical correlation analysis and dynamic time warping, and propose a method that jointly learns a feature space while aligning multiple time-sequences. The task is posed as a non-convex minimization problem, and a soft penalty is applied on the curvature of the warping path, in order to promote smooth warping paths. 

More recently, Kazlauskaite et al. \cite{kaz19p748} use a latent variable model that allows them to cluster multiple classes of observations, where inter-class observations are affected by unknown warpings.

\subsection*{Contribution}

The methods reviewed above consider a handful of observations, and rely on either computationally heavy optimizations or some approach that requires de-warping between pairs of signals, an operation prone to error, and which can be time-consuming. Consequently, they do not scale well with the number of observations. In contrast, we consider scenarios involving a large number of observations, and propose a very simple approach that does not require altering the observed sequences. This is our main point of departure from existing literature. We show that by considering a generalized `median' operation on the observations, one can actually recover the unknown template $f(\cdot)$ reliably. While the assumption on the availability of a large number of observations may not apply in every situation, we show how this can be feasible when data is streaming and contains identifiable features. 

\subsection*{Outline}
A general discussion that paves the way to the proposed method is provided in Section~\ref{sec:setup}. Following this, Section~\ref{sec:median} describes the proposed algorithm in detail. The algorithm is demonstrated with two experiments on ECG and PPG signals in Section~\ref{sec:experiment}. Finally, Section~\ref{sec:outlook} contains concluding remarks on how the idea can be extended to long signals, and images.

\section{Problem Setup}\label{sec:setup}

In this section, we discuss the problem setup, and provide some motivation for the main idea, to be further detailed in Section~\ref{sec:median}.
\renewcommand{\sc}{1}
\begin{figure}
\centering
\includegraphics[scale = \sc]{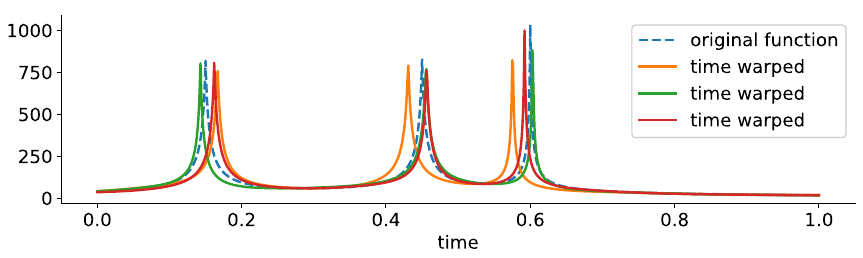}
\caption{An original function, and randomly warped observations, according to \eqref{eqn:f}. The function has three singularities at 0.15, 0.45, and 0.6. The location of the singularities is altered by warping.\label{fig:observations}}
\end{figure}

\begin{figure}
\centering
\includegraphics[scale = \sc]{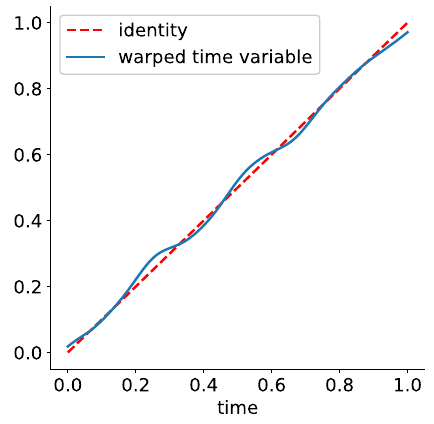}
\caption{A typical time-warping function used for producing the examples in Fig.~\ref{fig:observations}. \label{fig:warp}}
\end{figure}

For simplicity, we assume that $g_k$'s are of the form
\begin{equation}\label{eqn:gk}
g_k(t) = t + u_k(t),
\end{equation}
where $u_k(\cdot)$ is a sample from a zero mean stationary Gaussian process $U$. That is,
\begin{equation}\label{eqn:covariance}
\E(U(t_0), U(t_1)) = h(t_0 - t_1),
\end{equation}
for a positive definite kernel $h(t)$. For the synthetic experiments in this section, we use a squared exponential covariance function \cite{GPML}, which uses the kernel
\begin{equation}\label{eqn:kernel}
h(t) =  \alpha \exp( -  t^2/(2 \sigma^2)).
\end{equation}
This simple form does not rule out non-monotonic $g_k$'s but the probability of such curves is very small (see e.g., \cite{ram98p351} for an alternative formulation that respects monotonicity). 
\begin{remark}
The method does not depend on an intricate property of this particular Gaussian process model. The model allows us to develop some motivation in a concrete and objective manner.\qed
\end{remark}

For the setup outlined above, arguably the simplest estimate we can consider is the ensemble average. 
\begin{prop}\label{prop:average}
For $f_k(t)$, $g_k(t)$ as defined through \eqref{eqn:f}, \eqref{eqn:gk}, \eqref{eqn:covariance}, \eqref{eqn:kernel}, define the ensemble average as
\begin{equation}
\bar{f}_N(t) = \frac{1}{N}\,\sum_{k=1}^N\, f_k(t).
\end{equation}
Then, for all $t$, we have, almost surely,
\begin{equation}\label{eqn:conv}
\bar{f}(t) := \lim_{N \to \infty} \bar{f}_N(t) = (f \ast q)(t),
\end{equation}
provided the convolution exists, where $q(t)$ is a normal pdf with variance $ 1/h(0) = 1/\alpha$.
\begin{proof}
Note that for each $t$, $\{f_k(t)\}_k$ forms an iid sequence. So, by the strong law of large numbers, we have, for each $t$, $\bar{f}_N(t) \to \mathbb{E}(f_1(t))$ almost surely. But, since  $f_1(t) = f(g_1(t))$, we have,
\begin{align}
\mathbb{E}\bigl(f(g_1(t))\bigr) &= \int_u\,f(t + u)\,q(u)\,du \\&= \int_u\,f(t - u)\,q(u)\,du
\end{align}
where the second line follows by the symmetry of $q$.
\end{proof}
\end{prop}

\begin{figure}
\centering
\includegraphics[scale = \sc]{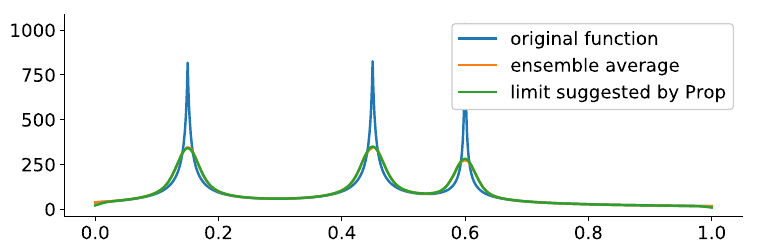}
\caption{The ensemble average leads to a smoothed estimate, as predicted via \eqref{eqn:conv} of Prop.~\ref{prop:average}. \label{fig:mean}}
\end{figure}

Prop.~\ref{prop:average} reveals when averaging can be effective or undesirable for this problem. If $f(\cdot)$ is known to be lowpass, (or Lipschitz with a small enough Lipschitz constant), it will undergo a less severe distortion due to blurring with the Gaussian kernel $q$ in \eqref{eqn:conv}. However, for $f(\cdot)$ with sharp features, $\bar{f}$ may not be a good estimate of $f$. Fig.~\ref{fig:mean} shows $f$ along with $\bar{f}_{1000}$ for the running synthetic example. 

Ignoring the smoothness of $f$ for a moment, one approach to estimating it could be to deconvolve $\bar{f}_N$ with $q$. In order for this to work, $\bar{f}_N$ needs to be very close to its limit value because deconvolution is a highly ill-conditioned problem. Unfortunately, for $N=1000$, the average still shows small deviations from the limit $\bar{f}$ for this example, which could be amplified by the ill-conditioned deconvolution operation.

For non-smooth functions, averaging is likely to blur interesting features, and may result in loss of information. But if the warping is not extreme, these features, even though misplaced, are preserved in every $f_k$. Following this observation, we restrict our estimate to be one of the observations, $\hat{f} \in \{f_k\}_k$. In a loose sense, this is similar to using the median as a robust estimate of the mean for when observations are scalar. However, in our problem, we do not have a natural ordering across $f_k$'s, so it is not clear how to define a median signal. Also, we do not know beforehand whether we can find an $f_k$ close enough to $f$. At this point, we can pose two questions : 
\begin{enumerate}[(i)]
\item\label{Q:i} [Existence] Given $\epsilon >0$, and some relevant norm $\| \cdot \|$, how many time-warped observations ($N$) do we need to ensure there exists $f_k$ with $\| f(t) - f_k(t) \| \leq \epsilon$?
\item \label{Q:ii} [Computing the center] Given a set of observations, how can we pick the observation $f_k$ that is closest to $f$?
\end{enumerate}

This letter is mainly concerned with the question about `computing the center', which is taken up in Section~\ref{sec:median}.  However, we make a few remarks on the question of `existence', to complete the discussion.
\subsection{Existence of a Feasible Solution}
According to the assumed probabilistic model on $f_k$'s, we need to compute the probability of the event $E_{N,\epsilon} = \{ \exists \, k \, \text{ s.t. } \| f - f_k\| \leq \epsilon\}$. But we cannot compute this probability because $f$ is unknown. Notice that, provided $f$ is smooth (Lipschitz) everywhere except for $K$ points, it is sufficient to ensure that $f_k$ is close to $f$ at those $K$ points. For our running example, there are three such points, $x \in \{0.15, 0.45, 0.6\}$. In order to ensure the existence of $f_k$ that is close to $f$, it suffices that $g_k(t)$ be sufficiently close to $t$ at those three points. We can therefore consider instead the probability of the event
\begin{equation}\label{eqn:Etilde}
\tilde{E}_{N,\delta} =  \{ \exists k \text{ s.t. } | g_k(t) - t | < \delta \text{ if $t$ is a singularity}\},
\end{equation}
for some $\delta$.

\begin{remark} 
We can, in principle, compute probabilities of events of the form
\begin{equation}
E'_{N, \delta} = \{ \exists \, k \, \text{ s.t. } \| t - g_k(t) \| \leq \delta \text{ for all } t\}.
\end{equation}
For a small $\delta$, this event ensures that the warped function will be close to the original function, and is less dependent on the properties of $f$ -- i.e., we do not need to know the locations of the singularities. Even though $E'_{N,\delta} \subset \tilde{E}_{N,\delta}$, for $f$ with a few singularities,  we have $P(\tilde{E}_{N,\delta}) \gg P(E'_{N, \delta})$. So this approach may give too conservative estimates.
\qed
\end{remark}

\begin{figure}
\centering
\includegraphics[scale = \sc]{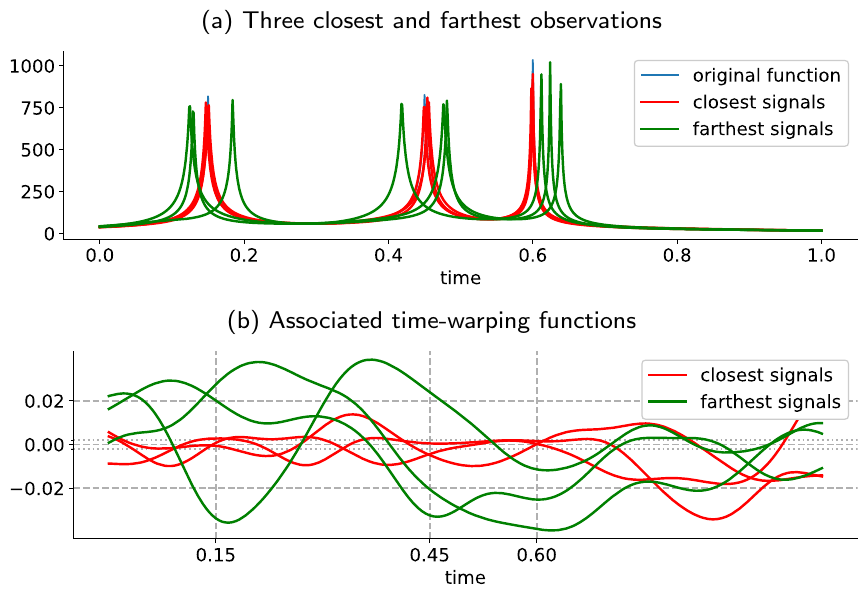}
\caption{a) The original function, along with the three closest and farthest observations, where distance is measured using an $\ell_2$ norm by an oracle. (b) Warping functions associated with the observations in (a).\label{fig:oracle}}
\end{figure}

In order to demonstrate this discussion, we conduct an experiment. We produce $N$ warped functions, and then rank these functions according to their $\ell_2$ distance to $f$. That is, $\tilde{f}_1$ is the closest to $f$ and $\tilde{f}_N$ is the farthest to $f$ in terms of $\ell_2$ distance. The top plot of Fig.~\ref{fig:oracle} shows the closest and farthest three observations (i.e., $\tilde{f}_1, \tilde{f}_2, \tilde{f}_3$ and $\tilde{f}_{N-2}, \tilde{f}_{N-1}, \tilde{f}_N$.). The bottom plot shows the warped time variables $g_1, g_2, g_3$ and $g_{N-2}, g_{N-1}, g_N$ that produced these functions. Notice that $g_1\dots g_3$ are not close to zero everywhere. But they approach zero near the singularities at 0.15, 0.45, 0.6.

The foregoing discussion outlines how we might answer Question~\eqref{Q:i} on the existence of a feasible solution. If we have information about the location or number of singularities of $f$, and how $g_i$'s are distributed, we can, in principle, compute an upper bound to ensure the existence of a suitable $f_k$. In practice, however, the many unknowns required in this procedure may prevent us to find a sharp bound.

In the following, we forgo Question~\eqref{Q:i}, and consider the more practical Question~\eqref{Q:ii}, on how to compute a center. We \text{assume} that a sufficient number of observations have been made.

\begin{figure}
\centering
\includegraphics[scale = \sc]{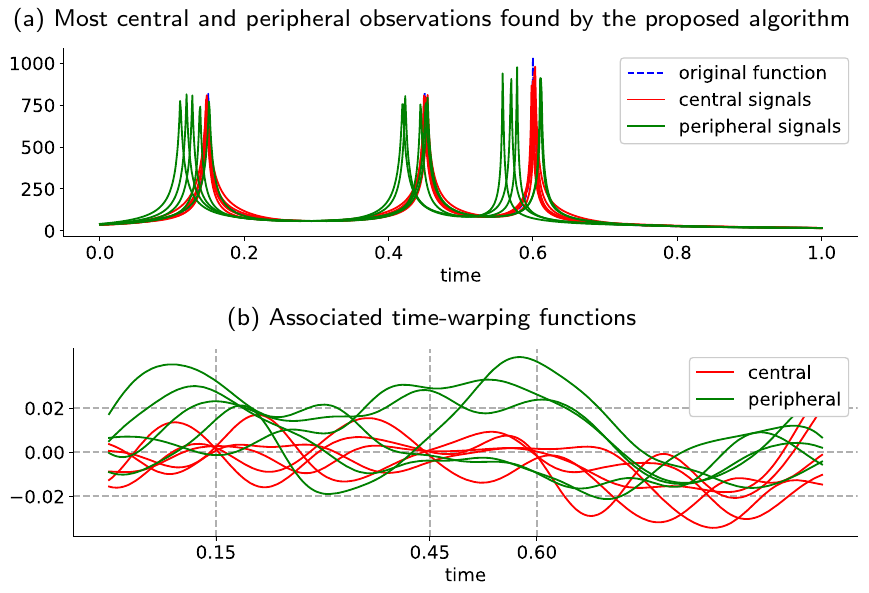}
\caption{(a) The five most central and most peripheral functions determined by the proposed algorithm. (b) Warping functions associated with the functions in (a). \label{fig:ordered}}
\end{figure}

\section{Finding the Median Signal}\label{sec:median}

Consider the manifold of signals obtained by warping a prototype signal $f(\cdot)$. Given samples from this manifold according to an unknown distribution, we would like to estimate $f(\cdot)$. Provided that warpings and distortions that $f(\cdot)$ goes through manifest some approximate symmetry, we expect the `center' of this (bounded) manifold to be close to $f(\cdot)$. To find this center, we consider the Dirichlet problem, which has a discrete counterpart for graphs \cite{DoyleSnell, gra06p768}.

\subsection{Heat Diffusion}
Suppose we have access to the bounded manifold. We randomly select two points on the manifold, and fix their `temperature' to 0 and 1. Assuming heat diffuses according to the manifold Laplacian (as explained below), we can, in principle, find the steady-state temperature at each point on the manifold. Intuitively, if a point $z$ lies `midway' between the selected points, it will have a temperature of $1/2$. Now consider repeating this procedure, each time selecting two seed points randomly, setting their temperatures to 0 and 1, and recording the steady state temperature of every other point in the manifold. Under this random experiment, for a given point in the manifold $z$, the steady state temperature of $z$, call it $T_z$, is a random variable, whose expected value is $1/2$ (shown below for the discrete case). However, if $z$ is close to the center, we expect the variance of $T_z$ to be small, because it will be roughly midway for many pairs of randomly selected seed points. We therefore declare $\var(T_z)$ to be a measure of centrality. The smaller $\var(T_z)$ is, the more central will be $z$.

\subsection{Discrete Dirichlet Problem}
In order to realize this with a finite set of observations, we consider the samples of the manifold to be the nodes of a weighted undirected graph. More concretely, let $f_i$, $f_j$ be two samples from this manifold. We set $w_{ij} = w_{ji} \geq 0$ to be an affinity measure (inverse of distance) between $f_i$ and $f_j$ (to be made precise). In this setting, let $x_{i}^k$, $x_j^k$ denote the heat distribution on the weighted graph at time instance $k$. If heat transfer occurs proportional to (a) affinity between nodes, (b) energy difference between nodes, then
\begin{equation}\label{eqn:heat}
x_i^{k+1} - x_i^k = \alpha\, \sum_j - w_{ij}\,(x_i^k - x_j^k),
\end{equation}
for some positive constant $\alpha$.
We can write this equation more compactly as $x^{k+1} - x^k = -\alpha\, L\,x^k$ using the 
`Laplacian matrix' $L$ \cite{Newman, Bollobas}, which is defined as,
\begin{equation}
L(i,j) = \begin{cases}
- w_{ij}, &\text{ if } i\neq j,\\
\sum_{k\neq i},  w_{ij}&\text{ if } i= j.
\end{cases}
\end{equation}
At steady state, we expect to have $L\,x^{\infty} = 0$. Now if we further impose $x_{i_0} = 0$, $x_{i_1} = 1$, for some $i_0$, $i_1$, the problem of determining the steady state temperature is equivalent to solving the following set of equations
\begin{equation}\label{eqn:dirichlet}
L\,x = 0, \text{ subject to }x_{i_0}= 0, \, x_{i_1} = 1.
\end{equation}
This problem is known as the (discrete) Dirichlet problem, and has a unique solution when the graph induced by $w_{ij}$ is connected  \cite{DoyleSnell, gra06p768}.

\begin{figure}
\centering
\includegraphics[scale = \sc]{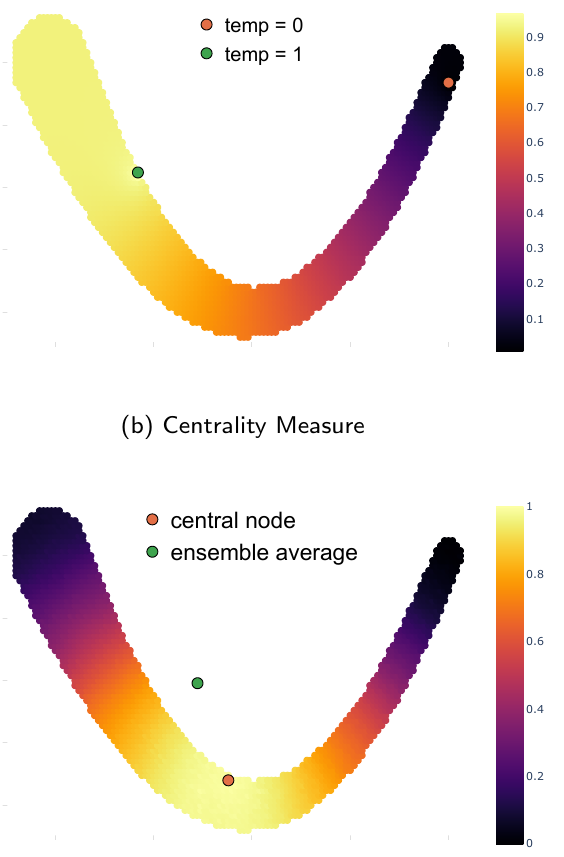}
\caption{The discrete Dirichlet Problem. (a) Two points are arbitrarily picked, and assigned temperatures of 0, and 1. The solution of the resulting Dirichlet problem assigns a temperature to the rest of the points. (b) The centrality measure obtained by repeating the above procedure and computing the temperature variance for each point. The figures shows the most `central' point, along with regular average.\label{fig:2D}}
\end{figure}

\begin{example}\label{ex:dirichlet}
To make the discussion more concrete, an example is shown in Fig.~\ref{fig:2D}a. We consider a bounded manifold in $\mathbb{R}^2$, in the shape of an asymmetric arc. We take 1833 points on this manifold, that lie on the intersection of the manifold and a regular rectangular grid. The Laplacian matrix is constructed by connecting each point to its north, south, east, west neighbors (with unit weights), if they exist. We select two points, assign them temperatures of 0, 1, and solve \eqref{eqn:dirichlet}. Fig.~\ref{fig:2D}a shows the resulting steady state heat distribution, where temperature is indicated by color. 
\qed
\end{example}

Consider now the experiment outlined in the example above, but suppose randomness is introduced by making the selection of the `seed' points random. That is, given $L$, suppose we \emph{randomly} select $i_0$, $i_1$, and solve \eqref{eqn:dirichlet}. Let $X$ denote the random vector that holds the resulting steady state temperature. Note that the number of components $X$ has is equal to the number of nodes of the graph.
\begin{prop}
$\mathbb{E}(X_k) = 1/2$ for all $k$.
\begin{proof}
Suppose the random experiment described above is performed so as to select $i_0$, $i_1$, and produce a realization of $X$, namely $x$. Define $z = \mathbf{1} - x$, where $\mathbf{1}$ denotes a vector of ones.
Then,
\begin{equation}
L\,z = L(\mathbf{1} - x) = -Lx  = 0,
\end{equation}
where we used the well-known property $L\,\mathbf{1} = 0$, which directly follows from the definition of the Laplacian matrix.
This observation implies that $z$ solves \eqref{eqn:dirichlet} if the roles of $i_0$ and $i_1$ are interchanged. Therefore, by the construction of the experiment, $z$ is observed with the same probability as $x$. Further, the average of $x$ and $z$ is $\frac{1}{2}\,\mathbf{1}$. Since $i_0$, $i_1$ were arbitrary, the claim follows by randomness of choosing $i_0$, $i_1$ in the experiment.
\end{proof}
\end{prop}
The proposition suggests that averaging the steady state temperature does not allow us to differentiate between the nodes of the graph. However, as discussed above, we do expect $X_k$ to have a tighter distribution around $1/2$ if node-$k$ is close to the `center' of the graph. We propose to use $\var(X_k)$ to be an inverse measure of centrality. This implicitly defines what we mean by the graph center, and leads to Algorithm~\ref{algo:center}.
\begin{algorithm}\caption{Computing a Measure of Centrality}\label{algo:center}
\begin{algorithmic}[1]
\Require Input graph Laplacian $L$ with $N$ nodes, number of trials $J$
\State Initialize the centrality measure $c$ to a zero vector 
\For{$j = 1:J$}
\State Randomly select $i, j \leq N$.
\State $x \gets \text{soln. of \eqref{eqn:dirichlet}}$
\State $c_k \gets c_k + (x_k  - 0.5)^2 / (J-1)$ for all $k \in \{1,2, \ldots N\}$
\EndFor
\end{algorithmic}
\end{algorithm}

\begin{example}[cont'd]
We run Algorithm~\ref{algo:center} with $J=1000$ trials. The resulting variance for a particular point gives an indication of how central the point is. Inverting this variance can be regarded as a measure of centrality. This centrality measure is displayed in Fig.~\ref{fig:2D}b. We also show the ensemble average (obtained by averaging the coordinates of each point), and the most `central' one, which maximizes the centrality measure. Notice that the ensemble average is outside the bounded set. In practice, this may mean that the ensemble average may not show the typical characteristics of the signal family under consideration. On the other hand, the `central' point obtained by the proposed method does lie in the bounded set, and agrees with our intuition that a center is a valid prototype for the considered set. \qed
\end{example}

\begin{remark}
The measure of centrality we describe above is a variation of an instance of `betweenness centrality'  for graphs \cite{Newman, new05p39}. There are other similar measures that aim to capture a similar feature for the nodes of a graph. Arguably the `best' choice ultimately depends on the application. For our purposes, we found the measure described above to be useful. However, we do not rule out the possibility that other measures of betweenness centrality may be useful in this context.
\qed
\end{remark}

\subsection{Complete Algorithm}
After clearing out some details, we will present the algorithm for estimating $f(\cdot)$ in this section.

\subsubsection*{Dealing with noise}
We propose to use the `center' of the observations as an estimate of $f(\cdot)$, which is defined to be the observation that has the maximum centrality measure, as computed by Alg.~\ref{algo:center}. However, if the observations are noisy, this means that our estimate will also be noisy. In order to reduce noise, we simply take the $K$ most central observations and compute their point-wise average. 

\subsubsection*{Affinity Measure}
As noted above, to define the Laplacian matrix, we need to compute an `affinity' $w_{ij}$ between any two observations $f_i$, $f_j$. As long as this $w_{ij}\geq0$, the resulting Laplacian matrix will be valid. In order to ensure that the Laplacian matrix describe a graph with a single component (an assumption we made in motivating the centrality measure), we can check the null-space of $L$, $\mathcal{N}(L)$. The graph is connected if and only if $\mathcal{N}(L)$ is one-dimensional. In that case, $\mathcal{N}(L)$ is spanned by the all-ones vector \cite{Bollobas}.

It is possible to make $L$ depend more on the shape of the underlying manifold, rather than the distribution of the samples on the manifold as discussed in Section 2.2 of \cite{laf06p784}. We followed that practice in our implementation. We also enforced the number of neighbors for each node to a user-specified value $M$. The resulting algorithm for constructing the Laplacian is given in Alg.~\ref{algo:Lap}. Note that the entries of the Laplacian matrix $L_{ij}$ obtained through this algorithm are equivalent to $-w_{ij}$, which act as a negative affinity measure.

\begin{algorithm}\caption{Constructing the Laplacian matrix}\label{algo:Lap}
\begin{algorithmic}[1]
\Require The observed vectors $f_1, \ldots, f_N$; Number of desired neighbors for each observation $M$
\State Compute  $d_{ij} = \| f_i - f_j \|_2^2$ for all $i\neq j$
\State $\tau_i \gets M^{\text{th}}$ smallest member of $\{d_{ij}\}_j$.
\State Set $\varepsilon \gets \text{median}(\tau_i)/3$.
\State $L_{ij} \gets - \exp( - d_{ij} / \varepsilon)$ for all $i\neq j$
\For{$i = 1:N$} \Comment Assert $M$ neighbors
\For{$j=1:i$}
\If {$d_{ij} > \tau_i$}
\State $L_{ij}  \gets 0$; $L_{ji} \gets 0$ \Comment Ensure symmetry
\EndIf
\EndFor
\EndFor
\State $d_i \gets \sum_{j\neq i} L_{ij}$ for all $i$ 
\State $L_{ij} \gets L_{ij} / (d_i\,d_j)$ for all $i\neq j$
\State $L_{ii} \gets -\sum_{j\neq i} L_{ji}$ \Comment Correct the diagonals
\end{algorithmic}
\end{algorithm}

We now have all of the ingredients to describe the complete algorithm.
The pseudo-code using these components is given in Alg.~\ref{algo:complete}.
\begin{algorithm}\caption{Estimating $f(\cdot)$}\label{algo:complete}
\begin{algorithmic}[1]
\Require The observed vectors $f_1, \ldots, f_N$, constant for Laplacian construction $M$, number of trials for computing the centrality measure $J$, number of central measure to use in averaging $K$
\State Compute the Laplacian matrix using $M$ as described in Alg.~\ref{algo:Lap}
\State Compute centrality measure using $J$ trials, as described in Alg.~\ref{algo:center}
\State Extract the $K$ most central observations, and compute their average as an estimate of $f$.
\end{algorithmic}
\end{algorithm}

\subsection{Note on Complexity}
The proposed algorithm works with Laplacian matrices, instead of signal samples. Therefore, at first sight, the complexity of the algorithm is determined by the size of the Laplacian matrix (or, the number of observations, $N$), rather than the length of each observation. To compute the proposed `centrality measure', the algorithm needs to solve linear system of equations involving the $N\times N$ Laplacian, $J$ times. Assuming a constant $J$, this may suggest a complexity on the order of $O(N^3)$ (which is an overstatement because we actually do not invert any matrices). But it is important to keep in mind the construction sets the number of neighbors for each node to  $M$. Therefore, the Laplacian is a sparse matrix, with $M+1$ non-zero entries in each row. With this restriction we aim to capture the dimension of the manifold where the warped signals live. For such a sparse system, the complexity of Gaussian elimination, under no additional assumptions, reduces to $O(N\,M^2)$. Therefore, the  complexity is related to the complexity of the warped-signal manifold. 

The preceding discussion does not specify how to select $J$ and $M$. Intuitively, one would expect $M$ to be related to the number of `anchor' points, or singularities in the signal. But in practice, we do not really know what $M$ should be, or how to set $J$ based on $M$ or $N$. Selection of these parameters affect performance as well as speed, but is a subject beyond the scope of the current letter.

\section{Experiments}\label{sec:experiment}
In the following, we demonstrate the feasibility of the proposed approach in two different experiments using real biomedical signals. Both experiments are on streams of data, and involve similar preprocessing stages. 

\subsection{Application on an Electrocardiogram Signal}

\begin{figure}
\centering
\includegraphics[scale = \sc]{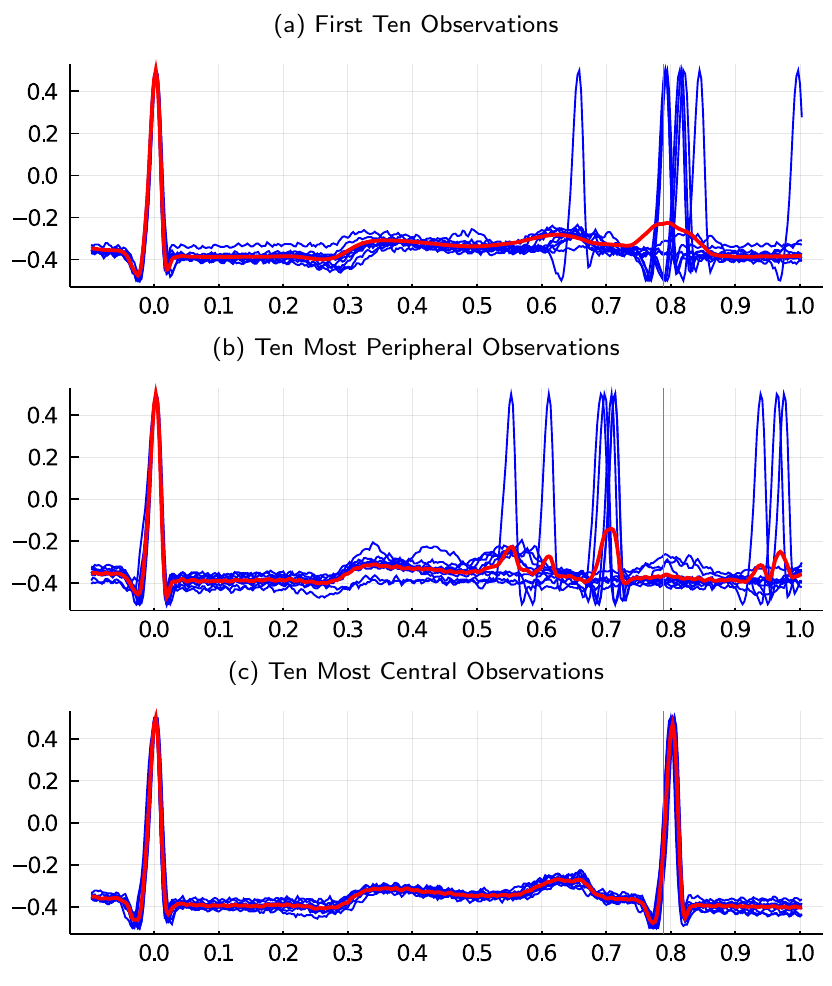}
\caption{Results from the experiment on an ECG time series. For each figure, the red curved is the average of the blue curves -- see the text for a description. \label{fig:ECG}}
\end{figure}

\begin{figure}[t!]
\centering
\includegraphics[scale = \sc]{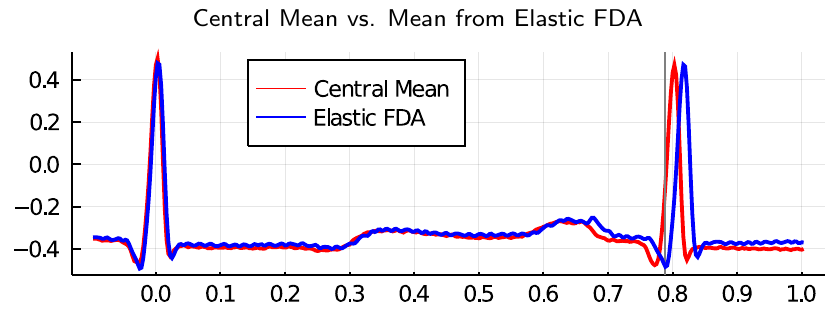}
\caption{Comparison of the proposed mean, computed using the central signals, and the mean produced by the Elastic FDA. The two signals look similar, but the computational cost for obtaining the proposed mean is much lower. \label{fig:ecg_compare}}
\end{figure}

We use a 15 minutes ECG recording sampled at 360 Hz, taken from the MIT-BIH arrhythmia database \cite{moo01p45}. 
We first extract the R-peaks of the 15-min long signal, obtaining $N = 1139$ peaks. For each peak, we define a signal excerpt $f_k$ by considering the interval $[t_k - 0.1, t_k + 1]$ (in seconds) where $t_k$ marks the location of the $k\thh$ peak. This gives us $N$=1139 signals $f_k$ with 397 samples each. The average heart rate is greater than 60 Hz, and a significant majority (if not all) of $f_k$'s contain more than one peak. Because of the construction, the first peak in each $f_k$ occurs at the same location, but the second peaks are not aligned. $f_1$, \ldots $f_{10}$ are shown in Fig.~\ref{fig:ECG}a. Also shown in red in the same figure is the average of all $N$ $f_k$'s. Notice that due to misalignment, the average of the second peak is very weak.

\begin{figure}[t!]
\centering
\includegraphics[scale = \sc]{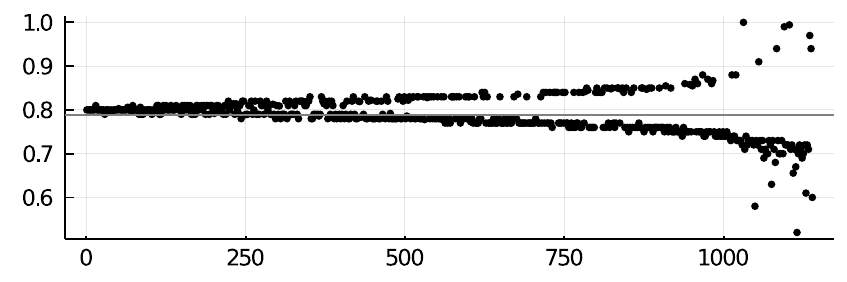}
\caption{Peak to peak distances with respect to the sorted indices. The algorithm places the observations where the peak to peak distances are too large or small to the periphery. \label{fig:P2P}}
\end{figure}

In order to find out the central and peripheral $f_k$'s, we apply our algorithm. The ten most peripheral signals, and their average is shown in Fig.~\ref{fig:ECG}b. Notice that the positions of the peaks vary by a lot. Consequently, their average is not very useful. The ten most central observations are shown in Fig.~\ref{fig:ECG}c. These are well-aligned, and return a coherent average. Finally, Fig.~\ref{fig:ecg_compare} shows the prototype signal obtained using the Julia ``Elastic FDA" package \cite{Tucker2014}, along with the average shown in Fig.~\ref{fig:ECG}c. Specifically, we used the function \texttt{srsf\_align} to align the curves. The running time of this function is considerably longer compared to the proposed algorithm. Because of this, we used only the first 100 $f_k$'s as input to the algorithm. We remark that both algorithms give meaningful averages, and we are \emph{not} claiming that the proposed algorithm's output is better than that of Elastic FDA. However, it is much \emph{faster} because it avoids the computationally heavy stage of curve alignment. Our main thesis is that such a stage is not necessary if we have enough time-warped observations.

The measure of centrality can also be used for anomaly detection in this example. For that, we sorted $f_k$ according to the centrality measure, so that $f_{p(i)}$ denotes the $i^{\text{th}}$ most central observation. We also compute the distance between the peaks in each $f_{k}$, and denoted the distance between the peaks in the observation $f_{p(i)}$ as $u(i)$. Fig.~\ref{fig:P2P} show the sequence of $u(i)$'s. We see that the proposed measure places observations with more common peak to peak distances in the center. This in turn allows to identify anomalous heart beats according to the long recording, even though they appear perfectly normal when viewed individually. For this example, this ordering could also be realized via an inspection of the peak to peak distances. But in other contexts where easily identifiable anchor points  are not present, such an approach might fail, while the proposed algorithm could still be used.

\subsection{Estimating a Normal PPG Waveform}\label{sec:PPG}

For this experiment, we use an 8-minute PPG recording taken from the database described in \cite{pim17p14}. Preprocessing is similar to that in the ECG experiment :  we first locate the peaks, and for each peak we consider the signal excerpt in the interval $[t - 1, t+1]$, where $t$ denotes the peak location. 

In contrast to ECG, PPG is a smooth signal. Also, the `average' waveform of PPG carries information about the condition of the subject \cite{elg12p14, all07}. But the waveforms go through various distortions from beat to beat. We employ the proposed algorithm to estimate the average waveform.

\begin{figure}
\centering
\includegraphics[scale = \sc]{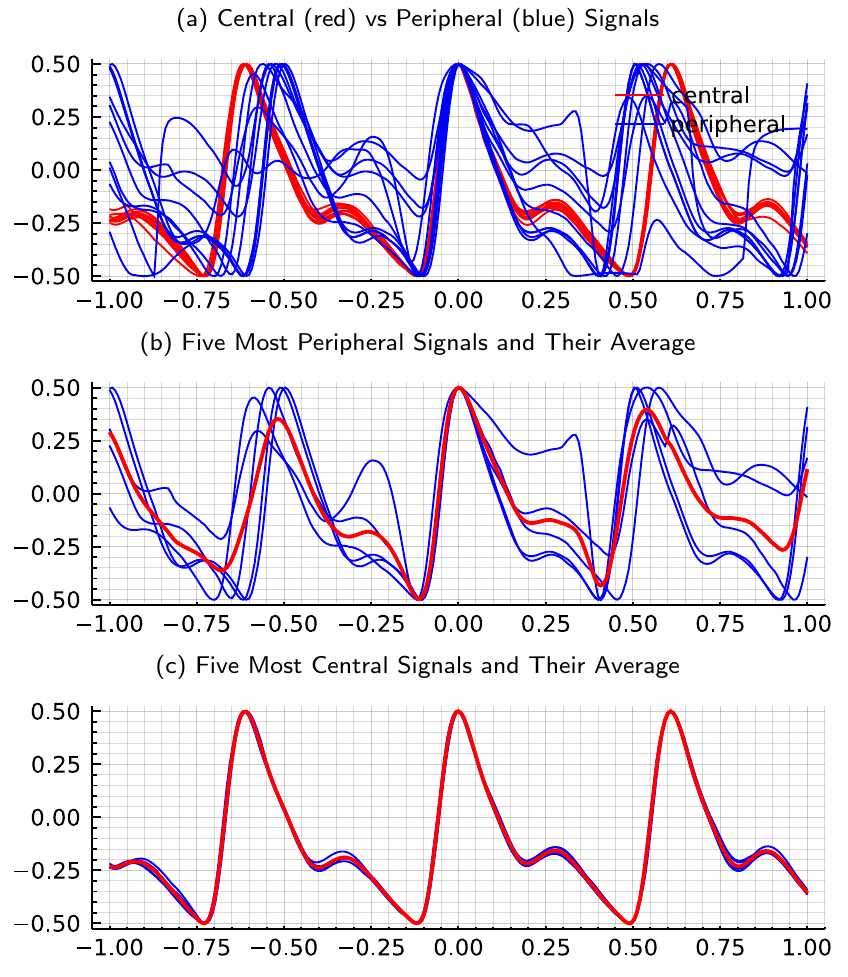}
\caption{Results from the experiment on a PPG time series. (a) Shows the ten most central (red), and peripheral (blue) signals, (b) shows the five most peripheral signals (blue) and their average (red), (c) shows the five most central signals (blue) and their average (red).  \label{fig:PPG}}
\end{figure}

\begin{figure}[t!]
\centering
\includegraphics[scale = \sc]{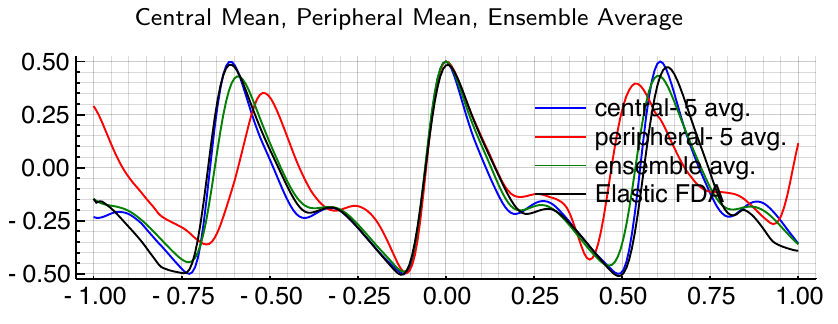}
\caption{Comparison of average signals, obtained using the 5 most central signals, 5 most peripheral signals, and the whole set of available signals.\label{fig:ppg_compare}}
\end{figure}

The ten most central and peripheral observations are shown in Fig.~\ref{fig:PPG}a. Using these, we compute peripheral and  central averages, as shown in Fig.~\ref{fig:PPG}b, c. The central signals are well-aligned, and the warping that relates one of these signals to another is very close to the identity. On the other hand, there is a significant variation between the peripheral signals. These signals lie on the outskirts of the manifold of warped signals. Finally, Fig.~\ref{fig:ppg_compare} shows the different averages, obtained by using the 5 most central, 5 most peripheral, the whole set of observations, and that obtained by elastic FDA \cite{Tucker2014}. Among these, the proposed central-5 average appears to preserve the details best, followed by the elastic FDA mean estimate. Notice that simply averaging the whole set of observations is not desirable because it also uses the peripheral signals, which show significant deviation.

\section{Outlook}\label{sec:outlook}

A basic assumption underlying the proposed method is that the number of singularities is small, compared to the number of observations, per the definition of the event $E_{N,\delta}$ in  \eqref{eqn:Etilde}. However, if the observed signals are long, this might disturb the balance between the number of singularities and the number of observations. In that case, we may no longer be able to find a single $f_k$ that is close to the original $f$. For such a scenario, one simple trick we found useful is to dissect the signal : we chunk the observations into overlapping pieces, ending up with $T$ pieces with $N$ observations. For each piece, we run the proposed algorithm, and find an estimate of $f$ for that particular piece. Then we can combine these estimates smoothly using windows with tapered ends. However, we should apply this trick with caution, because in the most extreme case where the chunk size is a single sample, we converge to a point-wise median, which we know is a poor estimate for the problem at hand.

Another extension of interest is images. In that case, we are likely to be in the regime described above -- too many singularity points relative to the number of observations. But we think that the outlined overlap-add approach would be feasible for that case. We hope to investigate application to images in future work.

\bibliographystyle{plain}


\begin{thebibliography}{10}

\bibitem{all07}
J.~Allen.
\newblock Photoplethysmography and its application in clinical physiological
  measurement.
\newblock {\em Physiological Measurement}, 28(3), 2007.

\bibitem{Bollobas}
B.~Bollob\'{a}s.
\newblock {\em Modern Graph Theory}.
\newblock Springer, 1998.

\bibitem{chen}
V.~C. Chen.
\newblock {\em Micro-Doppler Effect in Radar}.
\newblock Artech House, 2019.

\bibitem{DoyleSnell}
P.~G. Doyle and J.~L. Snell.
\newblock Random walks and electric networks, 2000.

\bibitem{elg12p14}
M.~Elgendi.
\newblock On the analysis of fingertip photoplethysmogram signals.
\newblock {\em Current Cardiology Reviews}, 8:14--25, 2012.

\bibitem{gra06p768}
L.~{Grady}.
\newblock Random walks for image segmentation.
\newblock {\em IEEE Transactions on Pattern Analysis and Machine Intelligence},
  28(11):1768--1783, 2006.

\bibitem{jam07p480}
G.~James.
\newblock Curve alignment by moments.
\newblock {\em Annals of Applied Statistics}, 1(2):480--501, 2007.

\bibitem{kaz19p748}
I.~Kazlauskaite, C.~H. Ek, and N.~Campbell.
\newblock Gaussian process latent variable alignment learning.
\newblock In K.~Chaudhuri and M.~Sugiyama, editors, {\em Proceedings of Machine
  Learning Research}, volume~89 of {\em Proceedings of Machine Learning
  Research}, pages 748--757. PMLR, 16--18 Apr 2019.

\bibitem{knei92p266}
A.~Kneip and T.~Gasser.
\newblock Statistical tools to analyze data representing a sample of curves.
\newblock {\em The Annals of Statistics}, 20(3):1266--1305, 1992.

\bibitem{kur11nips}
S.~A. Kurtek, A.~Srivastava, and W.~Wu.
\newblock Signal estimation under random time-warpings and nonlinear signal
  alignment.
\newblock In J.~Shawe-Taylor, R.~S. Zemel, P.~L. Bartlett, F.~Pereira, and
  K.~Q. Weinberger, editors, {\em Advances in Neural Information Processing
  Systems 24}, pages 675--683. Curran Associates, Inc., 2011.

\bibitem{laf06p784}
S.~{Lafon}, Y.~{Keller}, and R.~R. {Coifman}.
\newblock Data fusion and multicue data matching by diffusion maps.
\newblock {\em IEEE Transactions on Pattern Analysis and Machine Intelligence},
  28(11):1784--1797, 2006.

\bibitem{moo01p45}
G.~B. {Moody} and R.~G. {Mark}.
\newblock The impact of the {MIT-BIH} arrhythmia database.
\newblock {\em IEEE Engineering in Medicine and Biology Magazine},
  20(3):45--50, 2001.

\bibitem{Newman}
M.~Newman.
\newblock {\em Networks}.
\newblock Oxford University Press, $2^{\text{nd}}$ edition, 2018.

\bibitem{new05p39}
M.~E.~J. Newman.
\newblock A measure of betweenness centrality based on random walks.
\newblock {\em Social Networks}, 27(1):39 -- 54, 2005.

\bibitem{pim17p14}
M.~A.~F. Pimentel, A.~E.~W. Johnson, P.~H. Charlton, D.~Birrenkott, P.~J.
  Watkinson, L.~Tarassenko, and D.~A Clifton.
\newblock Toward a robust estimation of respiratory rate from pulse oximeters.
\newblock {\em IEEE Transactions on Biomedical Engineering}, 64(8):1914 --
  1923, 2017.

\bibitem{ram98p351}
J.~O. Ramsay and Xiaochun Li.
\newblock Curve registration.
\newblock {\em Journal of the Royal Statistical Society. Series B (Statistical
  Methodology)}, 60(2):351--363, 1998.

\bibitem{FDA}
J.~O. Ramsay and B.~W. Silverman.
\newblock {\em Functional Data Analysis}.
\newblock Springer, {S}econd edition, 2005.

\bibitem{GPML}
C.~E. Rasmussen and C.~K.~.I. Williams.
\newblock {\em Gaussian Processes for Machine Learning}.
\newblock The MIT Press, 2005.

\bibitem{Tucker2014}
J.~D. Tucker.
\newblock {\em Functional Component Analysis and Regression Using Elastic
  Methods}.
\newblock PhD thesis, Florida State University, 2014.

\bibitem{tuc13p50}
J.~D. Tucker, W.~Wu, and A.~Srivastava.
\newblock Generative models for functional data using phase and amplitude
  separation.
\newblock {\em Comput. Stat. Data Anal.}, 61(C):50–66, May 2013.

\bibitem{zho16p279}
F.~{Zhou} and F.~{De la Torre}.
\newblock Generalized canonical time warping.
\newblock {\em IEEE Transactions on Pattern Analysis and Machine Intelligence},
  38(2):279--294, 2016.

\end{thebibliography}

\end{document}